\documentclass{aa}
	\usepackage{amsfonts}
	\usepackage{amsmath}
	\usepackage[english]{babel}
	\usepackage{color}
	\usepackage[varg]{txfonts}	
	\usepackage[utf8]{inputenc}  
	\usepackage{natbib}
	\bibpunct{(}{)}{;}{a}{}{,} 
	\DeclareMathOperator{\Tr}{tr}

\title{Semi-blind Bayesian inference of CMB map and power spectrum}

\author{Flavien Vansyngel\inst{\ref{inst1}\ref{inst2}}
\and Benjamin D. Wandelt\inst{\ref{inst1}\ref{inst2}\ref{instbdw1}\ref{instbdw2}}
\and Jean-Fran\c{c}ois Cardoso\inst{\ref{inst1}\ref{instjfc1}\ref{instjfc2}}
\and Karim Benabed\inst{\ref{inst1}\ref{inst2}}
}

\institute{Sorbonne Universités, UPMC Univ Paris 06, UMR 7095, Institut d’Astrophysique de Paris, F-75014, Paris, France \label{inst1}
\and CNRS, UMR 7095, Institut d’Astrophysique de Paris, F-75014, Paris, France \label{inst2}
\and Department of Physics, University of Illinois at Urbana-Champaign, Urbana, IL 61801, USA \label{instbdw1}
\and Department of Astronomy, University of Illinois at Urbana-Champaign, Urbana, IL 61801, USA \label{instbdw2}
\and Laboratoire Traitement et Communication de l'Information, CNRS, UMR 5141 and Télécom ParisTech, 46 rue Barrault F-75634 Paris Cedex 13, France \label{instjfc1}
\and APC, AstroParticule et Cosmologie, Université Paris Diderot \label{instjfc2}
}


\abstract{
We present a new blind formulation of the Cosmic Microwave Background (CMB) inference problem. The approach relies on a phenomenological model of the multi-frequency microwave sky without the need for physical models of the individual components. For all-sky and high resolution data, it unifies parts of the analysis that have previously been treated separately, such as component separation and power spectrum inference. We describe an efficient sampling scheme that fully explores the component separation uncertainties on the inferred CMB products such as maps and/or power spectra. External information about individual components can be incorporated as a prior giving a flexible way to progressively and continuously introduce physical component separation from a maximally blind approach. We connect our Bayesian formalism to existing approaches such as Commander, SMICA and ILC, and discuss possible future extensions.
}

\begin{document}

\maketitle


\section{Introduction}

Observations of the Cosmic Microwave Background (CMB) constrain cosmological models. In particular, the CMB fluctuations are very sensitive to the parameters of the current standard model of cosmology \citep{Jungman:1995bz}. Current and future experiments designed for CMB analysis are signal dominated \citep{PlanckIII, Schaffer:2011mz, Bouchet:2011ck, Baumann:2008aq, Andre:2013nfa}. Therefore the remaining issue in deriving cosmological information from CMB is the separation between the CMB signal and the foregrounds signals. Being able to propagate component separation uncertainties to final constraints on fundamental physics is a leading issue in CMB analysis.

Most CMB experiments, such as the ongoing Planck mission \citep{PlanckI}, observe in the microwave domain. The CMB is not the only emission that is received when observing from the solar system at these frequencies. Free-free, synchrotron and thermal dust emissions emanating from our galaxy are among the most intense signals in the microwave domain \citep{Sehgal:2009xv}. An observation of the sky at these frequencies is therefore a mixture of the photons from the different sources. Therefore the CMB must be extracted through component separation techniques.

Apart from the assumption that the CMB emission law follows a black body, the method presented in the present paper makes no assumption about foreground emission. We make use of Independent Component Analysis (ICA) after assuming the mutual independence of the different signals constituting the data. Blind separation of independent sources (e.g. \cite{Cardoso:1998}) is a very general process that finds applications in various fields, from telecommunication to biomedical signals. Blind ICA has previously been applied in cosmology, particularly in CMB analysis. FastICA \citep{Baccigalupi:2000xy, Maino:2001vz} and SMICA \citep{2002astro.ph..9466C,2008arXiv0803.1814C,Delabrouille:2002kz} are two examples of that class of methods. Other methods, as GMCA \citep{Starck:896984,2013A&A...552A.133S}, exploit sparsity rather than independence to discern between different signals. In this paper we adopt the first approach and we propose a Bayesian instance of semi-blind ICA.

The different component separation methods are mainly characterised by two important aspects, the basis in which the data are expressed and the parametrisation of the data. Current methods performs the separation in different bases such as pixel space \citep{Eriksen:2005dr}, spherical harmonic space \citep{Tegmark:1997zz, Delabrouille:2002kz} or needlet space \citep{Delabrouille:2008qd, Moudden:2004wi, FernandezCobos:2011bm}. Their description of the data involves either a non-parametric model and exploits the independence between the CMB and the non-CMB component only -- e.g. NILC \citep{Delabrouille:2008qd}, SEVEM \citep{FernandezCobos:2011bm} -- or a parametric model that is fitted to the data -- e.g. Commander \citep{Eriksen:2005dr}. Intermediate between this these two, the SMICA method assumes coherence through frequency and complete independence of the components, fitting a non-parametric foreground model to the data \textit{via} likelihood maximisation.

One step forward is to chose a generic statistical model of the components based on generic assumptions (e.g. statistical independence of the components, spatial coherence between frequencies, spatial or angular scale statistical independence) which then allow a full Bayesian exploration of the posterior density. The introduction of a simple but full generative model that approximates the stochastic model of the component permits propagating the uncertainties within that model. The simplifying assumptions allow either a numerical marginalisation over all nuisance parameters \citep{2008arXiv0805.0093G} or, as in this paper, a full exploration of the model and a joint sampling of both the component maps and power spectra. The goal is to infer a CMB map and power spectrum, not to produce physical maps of the non-CMB components. 

This paper is organised as follows. The method and the simulations to which it is applied are described in Sect.~\ref{sec:method} and \ref{sec:sim}. Section~\ref{sec:results} presents the results. The robustness of the method is analysed in Sect.~\ref{sec:moch}. The method is compared to previous component separation methods in Sect.~\ref{sec:prevmeth}. We conclude and comment on future directions for development of these ideas in Sects.~\ref{sec:discussion} and \ref{sec:conclusion}.


\section{Method}
\label{sec:method}

\subsection{Data model}
We model the data as signal plus noise and the observed signal is assumed to be a linear mixture of several diffuse emissions. Hence the following decomposition for the piece of data $d_{ilm}$ contained in the spherical harmonic coefficient $(l,m)$ of the observation map at frequency $i$ (over $\mathrm{N_f}$ frequencies)
\begin{equation}
\label{eq:datamod}
d_{i\ell m} = \sum_{k=1}^{\mathrm{N_c}} A_{ik} s_{k\ell m} + n_{i\ell m} \, ,
\end{equation}
where the sum runs over the assumed number of components $\mathrm{N_c}$, $s_k=\left\lbrace s_{k\ell m} ; \ell=\ell_{\mathrm{min}}\dotso \ell_{\mathrm{max}}, m=-\ell\dotso \ell \right\rbrace $ is the spherical harmonic transform of the $k$th component map, $A_{ik}$ is the amount of component $k$ at frequency $i$ and $n_{i\ell m}$ is the amount of instrumental noise present in $d_{i\ell m}$. We suppose the data to be beam-corrected since debeaming in harmonic space is performed by just dividing the data by the beam transfer functions. Eq.~\ref{eq:datamod} reads in matrix form
\begin{equation}
\label{eq:datamod2}
d_{\ell m} = As_{\ell m} + n_{\ell m} \, .
\end{equation}
The $\mathrm{N_c}\times\mathrm{N_f}$ matrix $A$, which gathers all the coefficients $A_{ik}$, is called the \textit{mixing matrix}. The noise is assumed to be Gaussian with zero mean and $\langle n_{i\ell m}n_{i^\prime\ell^\prime m^\prime}\rangle = \nu^2_{i\ell}\delta_{ii^\prime}\delta_{\ell\ell^\prime}\delta_{mm^\prime}$. We use $C$ to denote the power spectra of the components and $\langle s_{k\ell m}s_{k^\prime\ell^\prime m^\prime}\rangle = C_{k\ell}\delta_{kk^\prime}\delta_{\ell\ell^\prime}\delta_{mm^\prime}$. The covariance of the data predicted by the model of Eq.~\ref{eq:datamod2} is then
\begin{equation}
\label{eq:modelcov}
R_\ell = AC_\ell A^T+N_\ell \, ,
\end{equation}
where $\left(C_\ell\right)_{kk^\prime}=\delta_{kk^\prime}C_{k\ell}$ and $\left(N_\ell\right)_{ii^\prime}=\delta_{ii^\prime}\nu^2_{i\ell}$ is the noise covariance at multipole~$\ell$.

The contribution from point sources is neglected in this work, but is discussed in Sect.~\ref{sec:discussion}.

\subsection{The SMICA likelihood}
SMICA \citep{2008arXiv0803.1814C} is a blind method working in harmonic space that provides an estimate of auto and cross power spectra and frequency spectra of the various components in the data. The parameters are determined by fitting the empirical spectral covariance of the data to the covariance of the model of Eq.~\ref{eq:modelcov}. This procedure is equivalent to the maximisation of the SMICA likelihood $\mathcal{L}_\text{SMICA}$ that is obtained after assuming the independence and the Gaussianity of the components, and
\begin{align}
\label{eq:smica}
-2 \log \mathcal{L}_\text{SMICA}\left(d \,\vert\,A,C\right) = \sum_{\ell,m} \log\left| 2\pi R_{\ell} \right| + d_{\ell m}^T R_{\ell}^{-1} d_{\ell m} \, .
\end{align}

Links between SMICA and the present method are detailed in Sect.~\ref{sec:prevmeth}.

\subsection{Bayesian formulation}
The aim of the method is to provide a joint probability distribution over the mixing matrix $A$,  the component maps $s$ and component power spectra $\left(A,s,C\right)$, knowing the data. If desired, prior knowledge on the parameters can be added. This formulation can be quantitatively written using Bayes theorem
\begin{equation}
\label{eq:bayesTh}
P\left(A,s,C\,\vert\,d\right) \propto \mathcal{L}\left(d\,\vert\,A,s\right) P\left(A,s,C\right) \, .
\end{equation}
The likelihood $\mathcal{L}\left(d\,\vert\,A,s\right)$ encodes the model chosen together with instrumental properties and the prior distribution $P\left(A,s,C\right)$ encodes the information already available about $A$, $s$, and $C$.
The power spectra are hyperparameters of the model since they parametrize the prior distribution on the component $s$.

Considering the power spectra as additional stochastic parameters has two motivations. First, it introduces more flexibility in the modelling of the components. Second and importantly, the posterior provides an inference of the power spectra, jointly with the mixing matrix and the component maps. Thus, errors introduced by the component separation step are propagated to the power spectra.

\subsubsection{The likelihood and marginalisation}
If $A$ and $s_{\ell m}$ in Eq.~\ref{eq:datamod2} are kept fixed then the stochasticity of $d_{\ell m}$ relies on the instrumental noise $n_{\ell m}$ only. Noise properties of an instrument are usually well determined but complex. Considering the noise to be Gaussian, independent from frequency to frequency is a good approximation for the WMAP and Planck missions \citep{PlanckIII,2011ApJS..192...14J}. Thus the Probability Distribution Function (PDF) for the noise is taken to be a multivariate normal distribution with zero mean and covariance $N$. The noise is supposed to be stationary, which imply that $N$ is diagonal in harmonic space. Therefore the likelihood is a product of Gaussians with mean $As_{\ell m}$ and covariance $N_\ell$, i.e. $\mathcal{L}(d\,\vert\,A,s) = \prod_{\ell m} \mathcal{N}(d_{\ell m};As_{\ell m},N_\ell)$.

If one wants to recover high resolution maps, then the posterior PDF is defined on a $\sim\!\! 10^7$-dimension parameter space. The usual sampling schemes used to draw from intractable PDF's are then inefficient because of high correlation length in the chains. One solution is to split the sampling problem into several sampling problems:
\begin{equation}
\label{eq:facto}
P(A,s,C\,\vert\,d) = P(s\,\vert\,A,C,d) P(A,C\,\vert\,d) \, .
\end{equation}
Sampling $\lbrace A,C\rbrace$ from the marginal $P(A,C\,\vert\,d)$ and then post-processing these samples to sample the component maps $s$ is equivalent to sampling $\lbrace A,C \rbrace$ from $P\left(A,s,C\,\vert\,d\right)$. The marginal is defined on a parameter space with a considerably reduced number of dimensions, the number of parameters being of the order of $10^3$, and is therefore simpler to sample from. Besides, sampling the maps is much more time-consuming than sampling the mixing matrix and the covariance.

By using the definition of a marginal distribution and the Bayes theorem of Eq.~\ref{eq:bayesTh}
\begin{equation}
P(A,C\,\vert\,d) \propto \int \text{d}s \, \mathcal{L}(d\,\vert\,A,s) P(A,s,C) \, .
\end{equation}
If the $A$ and $\lbrace s,C\rbrace$ are independent in the prior distribution, and if the prior on $s$ is taken to be a Gaussian with zero mean and covariance $C$, then
\begin{equation}
\label{eq:margin}
P\left(A,C\,\vert\,d\right) \propto \mathcal{L}_\textsc{SMICA}(d \,\vert\, A,C) P(A)P(C) \, .
\end{equation}
This marginalisation implicitly assumes that $C$ and $N$ are diagonal in the same basis. Thus, exploring the marginal posterior amounts to explore the SMICA likelihood weighted by prior distributions on the mixing matrix and the power spectra. 

\subsubsection{Choice of priors}
\label{sec:prior}
As stated above, the mixing matrix $A$ and the components $s$ are independent in the prior PDF, i.e. $P(A,s,C)=P(A)P(s\,\vert\,C)P(C)$. In this section we describe and justify the choice of the prior distributions of our analysis. Prior information must be chosen with care as it can introduce biases. We opted for non-informative or mildly informative priors in order to keep the analysis as blind as possible, although the Bayesian formalism does allow inclusion of further information.

\paragraph{Component maps} As a prior on component maps, we put a zero mean Gaussian PDF uncorrelated from multipole to multipole and from component to component that is parametrized by its diagonal covariance $C$.

The simplest inflationary theories predict that CMB fluctuations are very nearly a Gaussian random field (e.g. \cite{Mukhanov:2013tua}). Therefore, choosing a Gaussian random field as a prior for the CMB is physically well motivated. The choice of a gaussian prior is for computational reasons. We will assess the impact of this approximation in Sect.~\ref{sec:moch}.

\paragraph{Mixing matrix} The spectral behaviour of the CMB is supposed to be perfectly known and to follow a black body law, which has the consequence for the CMB signal to have constant response through frequencies in the data. Thus, when expressed in thermodynamical units, the elements of the column related to the CMB are fixed to the same unit constant. Regarding the other columns, the elements are normalised with regard to a reference frequency. The normalisation is necessary to break a continuous degeneracy between each column of the mixing matrix and the corresponding component spectrum. Indeed, the mixing coefficients, i.e. how much a component is present in the data, and the standard deviation of this component can be chosen up to an arbitrary factor. In order to fix this degree of freedom, one element of each column of the mixing matrix is fixed to unity. A flat prior is applied on the remaining elements, such that $P(A) \propto 1$.

\paragraph{Component power spectra} All components are uncorrelated \textit{a priori}. For all multipole $\ell$, $C_\ell$ is diagonal and thus contains the power spectra of the components on the diagonal. The non-informative Jeffreys prior is applied on the power spectra, i.e. $P(C) \propto \prod_{k,\ell} 1/C_{k\ell}$.

\subsection{Sampling techniques}
The marginal posterior in Eq.~\ref{eq:margin} can not be sampled from directly and we need to explore the parameter space \textit{via} sampling. We adopt the Metropolis-Hastings formalism to draw samples of $\lbrace A, C \rbrace$ and then estimate the marginal posterior. Then the full posterior over $\lbrace A,s,C \rbrace$ is recovered sampling conditionally on $\lbrace A,C \rbrace$ as in Eq.~\ref{eq:facto}.

In practice we let the Metropolis-Hastings sampler evolves until it converges and draws enough (say $\mathrm{N_{sam}}$) uncorrelated samples of $\lbrace A,C \rbrace$. The chain $\{\lbrace A_n, C_n\rbrace\,;n=1\dots\mathrm{N_{sam}}\}$ provides an estimation of the marginal posterior. Then, for each sample $\lbrace A_n, C_n\rbrace$, we draw a sample of the component maps $s_n$ from the conditional $P(s\,\vert\,A\!\!\! =\!\!\! A_n, C\!\!\! =\!\!\! C_n,d)$. The chain $\lbrace \lbrace A_n, s_n, C_n \rbrace\,; n=1\dots\mathrm{N_{sam}}\rbrace$ provides an estimate of the full posterior $P(A,s,C\,\vert\,d)$.

The sampling of the $s_n$ is straightforward since the conditional $P(s\,\vert\,A,C,d)$ is a Gaussian distribution, independent from one harmonic coefficient to another. For each piece of maps $s_{\ell m}$ the mean and covariance are
\begin{eqnarray}
\mu_{\ell m} & = & \Sigma_\ell A^T N_\ell^{-1} d_{\ell m} \label{eq:mulm} \\ 
\Sigma_\ell & = & \left(A^T N_\ell^{-1} A + C_\ell^{-1}\right)^{-1} \, . \label{eq:sigmal}
\end{eqnarray}
Thus, the $\mu_{\ell m}$'s are obtained by Wiener-filtering the data. There is a loss of power at high multipole in the mean due to the filter. Sampling the maps corrects this and the samples of the maps have the correct covariance. Following \cite{PhysRevD.70.083511}, the sampling of the maps is done by solving the system
\[ \Sigma_\ell^{-1} s_{\ell m} = A^T N_\ell^{-1} d_{\ell m} + \xi_{\ell m} \]
where $\xi_{\ell m}$ is i.i.d. for each $s_{\ell m}$ from a Gaussian distribution with zero mean and covariance $\Sigma_\ell^{-1}$.

\subsection{The explicit posterior}

Given the model of Eq.~\ref{eq:datamod2} and since we have chosen a flat prior for the mixing matrix $A$ and a Gaussian prior for the component maps $s$ and the Jeffreys prior for the power spectra $C$, the full posterior we want to sample from is
\begin{flalign}
\label{eq:fullpost}
 & P(A, s ,C\,\vert\,d) \propto & \nonumber \\
 & \quad \prod_{\ell ,m} \left|N_\ell\right|^{-\frac{1}{2}} \exp-\frac{1}{2} \left(d_{\ell m} - As_{\ell m} \right)^T N_\ell^{-1} \left(d_{\ell m} - As_{\ell m} \right) & \nonumber \\
 & \quad \prod_{\ell ,m} \left|C_\ell\right|^{-\frac{1}{2}} \exp-\frac{1}{2} s_{\ell m}^T C_\ell^{-1} s_{\ell m} & \\
 & \quad \prod_{k,\ell} C_{k\ell}^{-1} \, . & \nonumber
\end{flalign}
In order to draw sample from the full posterior, we split the problem, see Eq.~\ref{eq:facto} where, in our case,
\begin{flalign}
 & P(A,C \,\vert\,d) \propto & \nonumber \\ & \quad \prod_\ell \left| R_\ell \right|^{-\frac{2\ell +1}{2}} \exp-\frac{1}{2}\Tr \left((2\ell+1) R_\ell^{-1} V_\ell\right) \prod_{k,\ell} C_{k\ell}^{-1} &  \label{eq:pac} \\
 & P(s\,\vert\,A,C,d) \propto & \nonumber \\ & \quad \prod_{\ell ,m} \left|\Sigma_\ell\right|^{-\frac{1}{2}}\exp-\frac{1}{2}\left(s_{\ell m}- \mu_{\ell m}\right)^T \Sigma_\ell^{-1} \left(s_{\ell m} - \mu_{\ell m}\right) & \label{eq:psiac}
\end{flalign}
with $R_\ell$, $\mu_{\ell m}$ and $\Sigma_\ell$ respectively defined in Eqs.~\ref{eq:modelcov}, \ref{eq:mulm} and \ref{eq:sigmal}, and $V_\ell = \frac{1}{2\ell+1}\sum_m d_{\ell m}d_{\ell m}^T $.

The method consists in sampling $\lbrace A,C \rbrace$ from Eq.~\ref{eq:pac} using Metropolis-Hastings sampling and then choosing a subset of uncorrelated samples of $\lbrace A,C \rbrace$ to conditionally draw $s$ from the Gaussian in Eq.~\ref{eq:psiac}. The factorisation in Eq.~\ref{eq:facto} assures that this process amounts to sample from the full posterior of Eq.~\ref{eq:fullpost}.


\section{Simulations}
\label{sec:sim}

\begin{figure*}
\centerline{\includegraphics[scale=0.3]{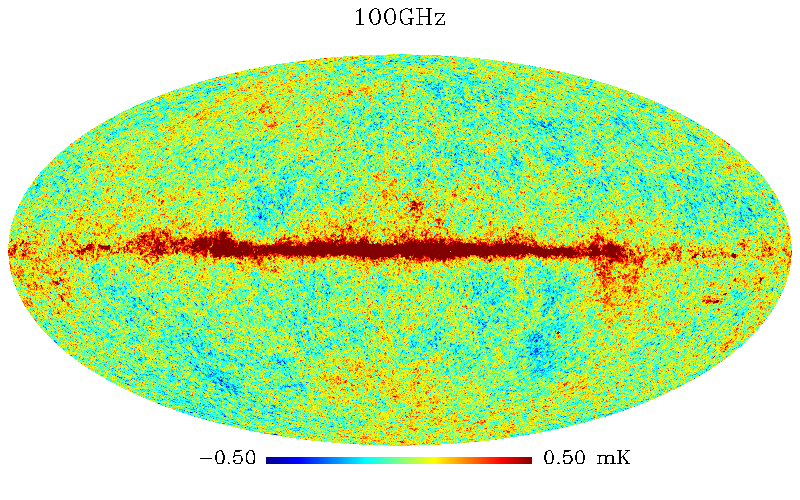}\includegraphics[scale=0.3]{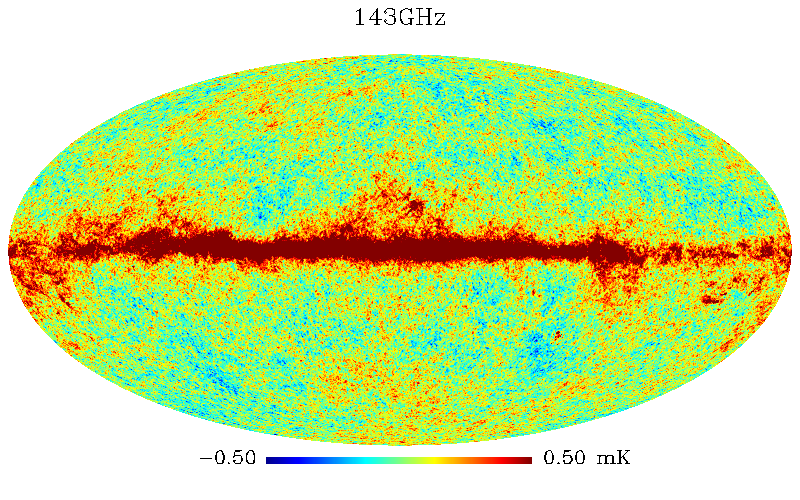}}
\centerline{\includegraphics[scale=0.3]{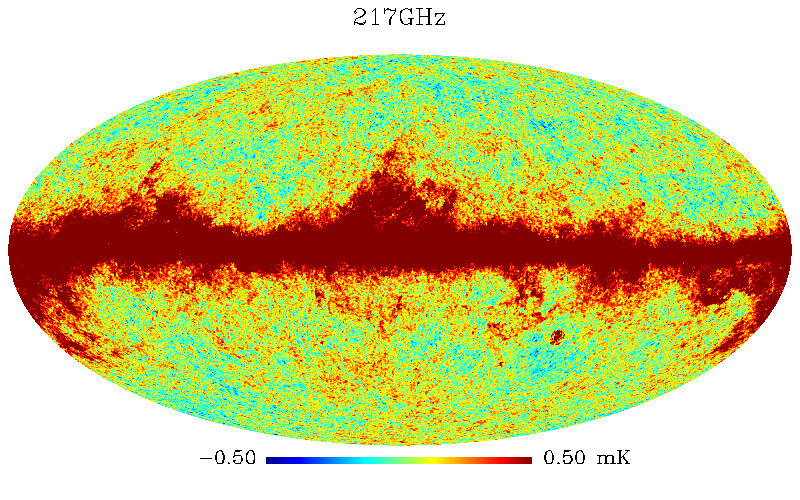}\includegraphics[scale=0.3]{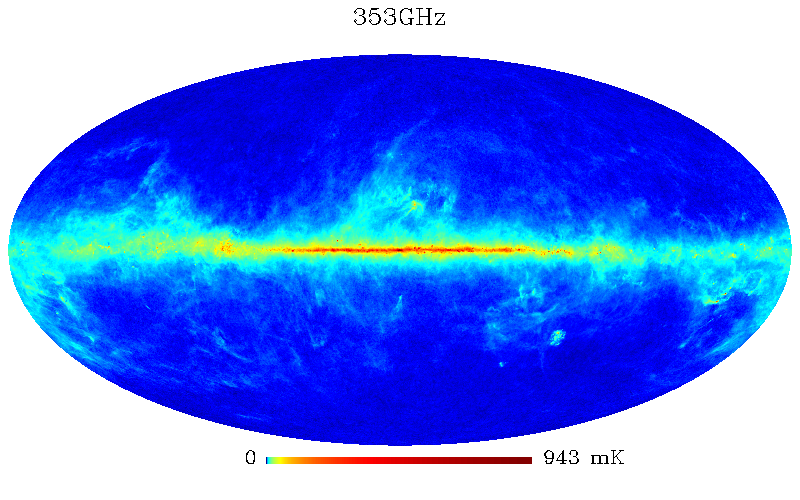}}
\centerline{\includegraphics[scale=0.5]{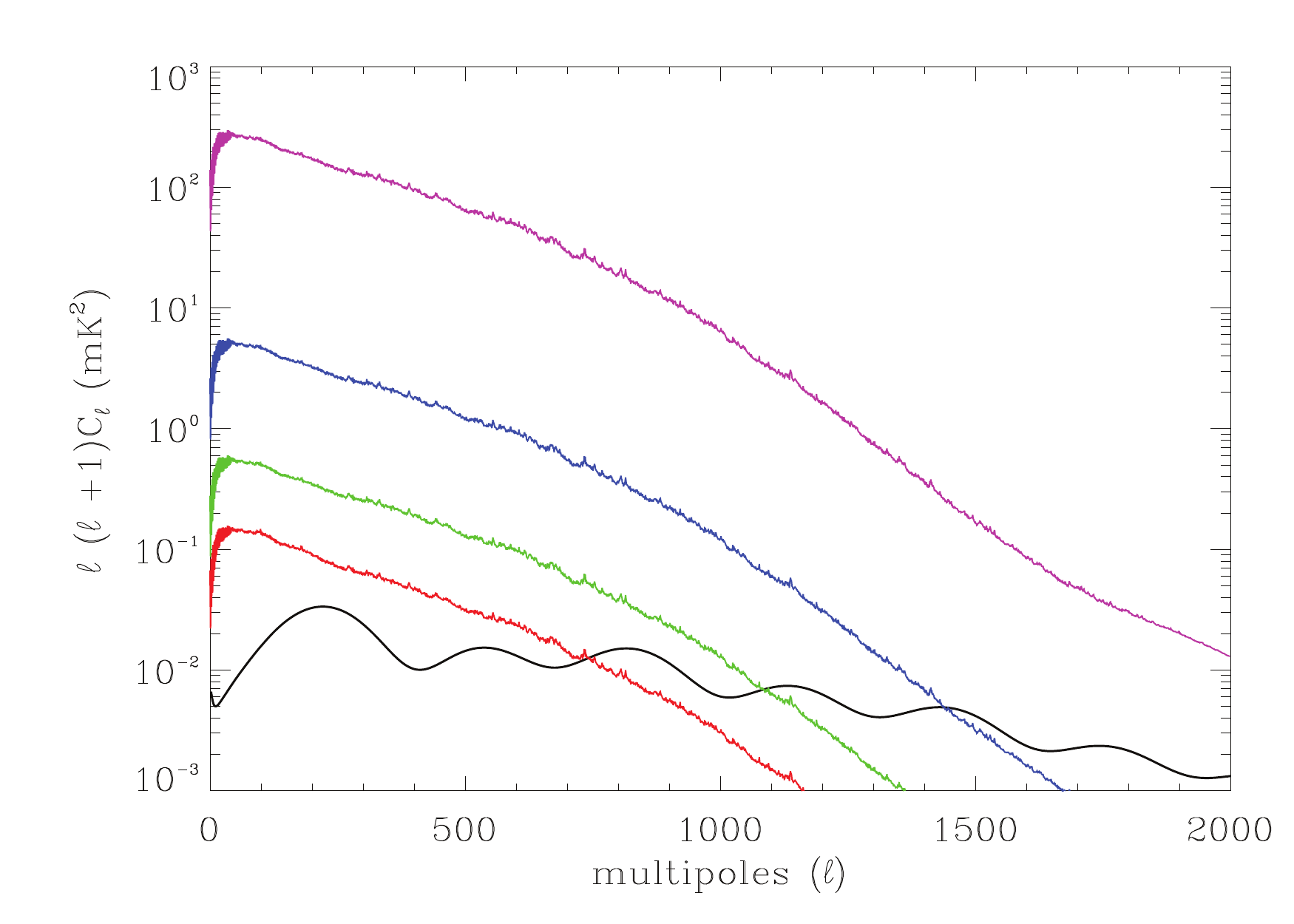}}
\caption{The simulated data maps at four of the Planck HFI frequencies, from 100GHz to 353GHz, using realistic spatial distributions of free-free and thermal dust emissions from the PSM. In these simulated data maps the templates of the component maps are scaled through frequency according to the mixing matrix. We chose to work with this set of channels because the CMB is the least contaminated by foregrounds and noise in this frequency range. The plot shows the power spectrum of the CMB (black line) and the level of foregrounds at each frequency channel in color (red to purple is 100GHz to 353GHz).}\label{fig:dmap}
\end{figure*}

For the purpose of this paper, we consider $\mathrm{N_c}=3$ components to be separated, observed at $\mathrm{N_f}=4$ HFI frequency channels, 100GHz, 143GHz 217GHz and 353GHz. The data maps are simulated according to the linear model of Eq.~\ref{eq:datamod2}. The set of simulations is a noisy composite of CMB, thermal dust emission and free-free emission.

Fig.~\ref{fig:dmap} shows the four simulated observation maps. The plot shows the CMB power spectrum in black and the power spectra of all the foregrounds at each frequency channel in color, from red to purple being from 100GHz to 353GHz.

\subsection{The component maps and their power spectra}
The data model of Eq.~\ref{eq:datamod2} implies that the components are coherent through frequencies. Therefore, one map only of each component is simulated.

The CMB map is simulated using the HEALPix software\footnote{\url{http://healpix.sourceforge.net}} \citep{2005ApJ...622..759G} from a power spectrum computed by the CAMB software \citep{Lewis:1999bs} in a standard $\Lambda$CDM model.

The spatial distributions of the foregrounds are simulated using the publicly available version of the Planck Sky Model (PSM) \citep{Delabrouille:2012ye}. The free-free map from the PSM has an electron temperature of 7000K, a power law with spectral index close to -0.15 and is a composite of maps from \citet{2003MNRAS.341..369D} and from the WMAP MEM map \citep{2012arXiv1212.5225B}. The thermal dust map from the PSM is simulated from the Schlegel-Finkbeiner-Davies map, from which the ultra-compact \textsc{HII} regions are subtracted. More details can be found in \citet{Delabrouille:2012ye}.

\subsection{The mixing matrix}
The CMB column of the mixing matrix is fixed to 1, and the normalising elements are also fixed to 1. The other elements agree with the PSM.

\subsection{The noise maps}
The noise is simulated at the map level, and is uncorrelated from pixel to pixel. The noise standard deviation maps are designed to be consistent with the scanning strategy chosen for the Planck spacecraft and is therefore anisotropic. In the harmonic domain the noise is characterized by one white power spectrum per frequency, derived from the noise standard deviation maps. Our inference approximate the noise as isotropic. The impact of this approximation will be assessed in Sect.~\ref{sec:moch}.


\section{Joint inference of CMB map and power spectrum}
\label{sec:results}

\begin{figure}
\centerline{\includegraphics[scale=0.5]{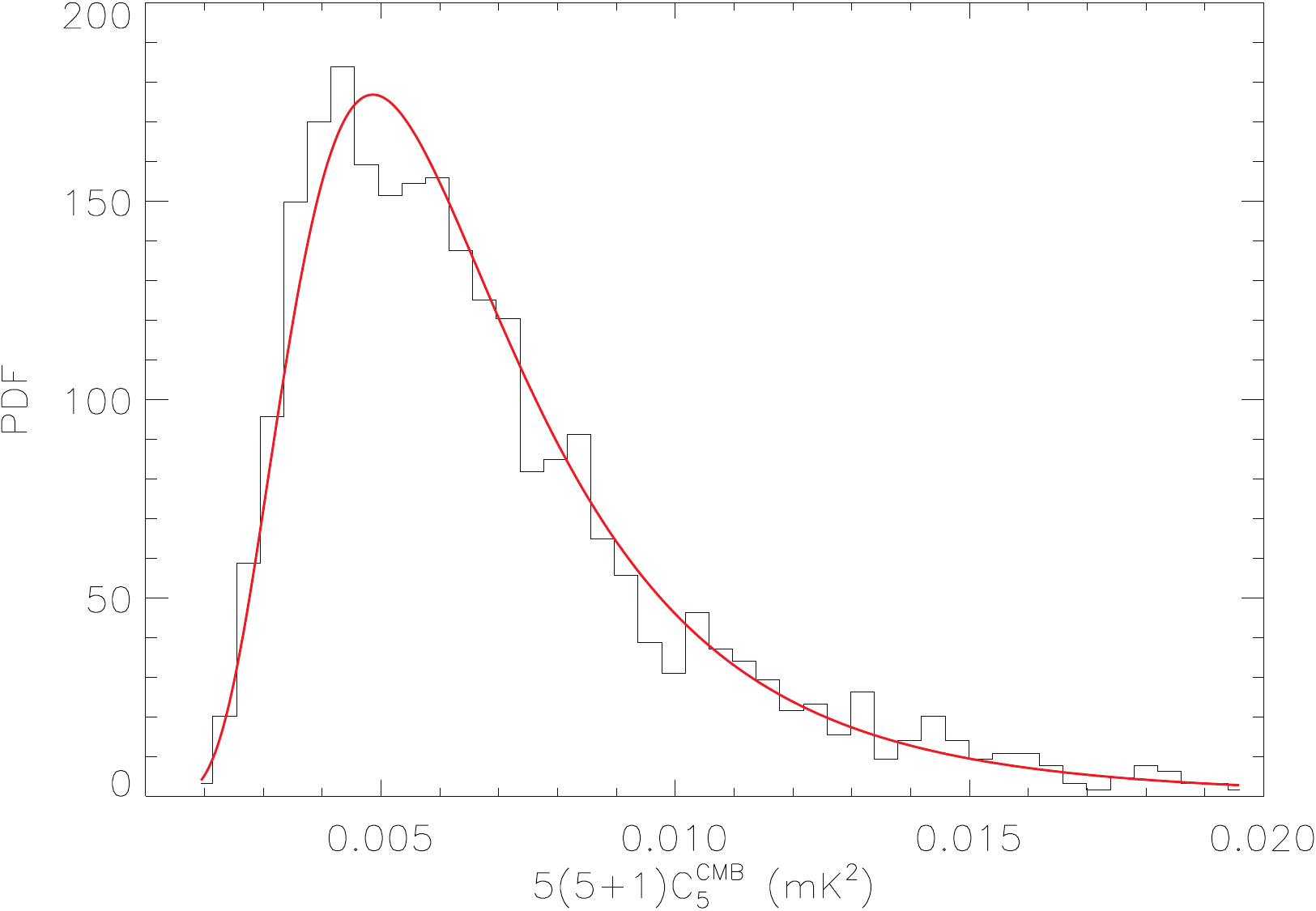}}
\caption{Posterior PDF marginalised over all parameters but the CMB power spectrum at multipole $\ell=5$. The histogram is an estimation of the PDF and the solid red curve is the best fit of an inverse-gamma function to the histogram.} \label{fig:fit}
\end{figure}

\begin{figure}
\centerline{\includegraphics[scale=0.5]{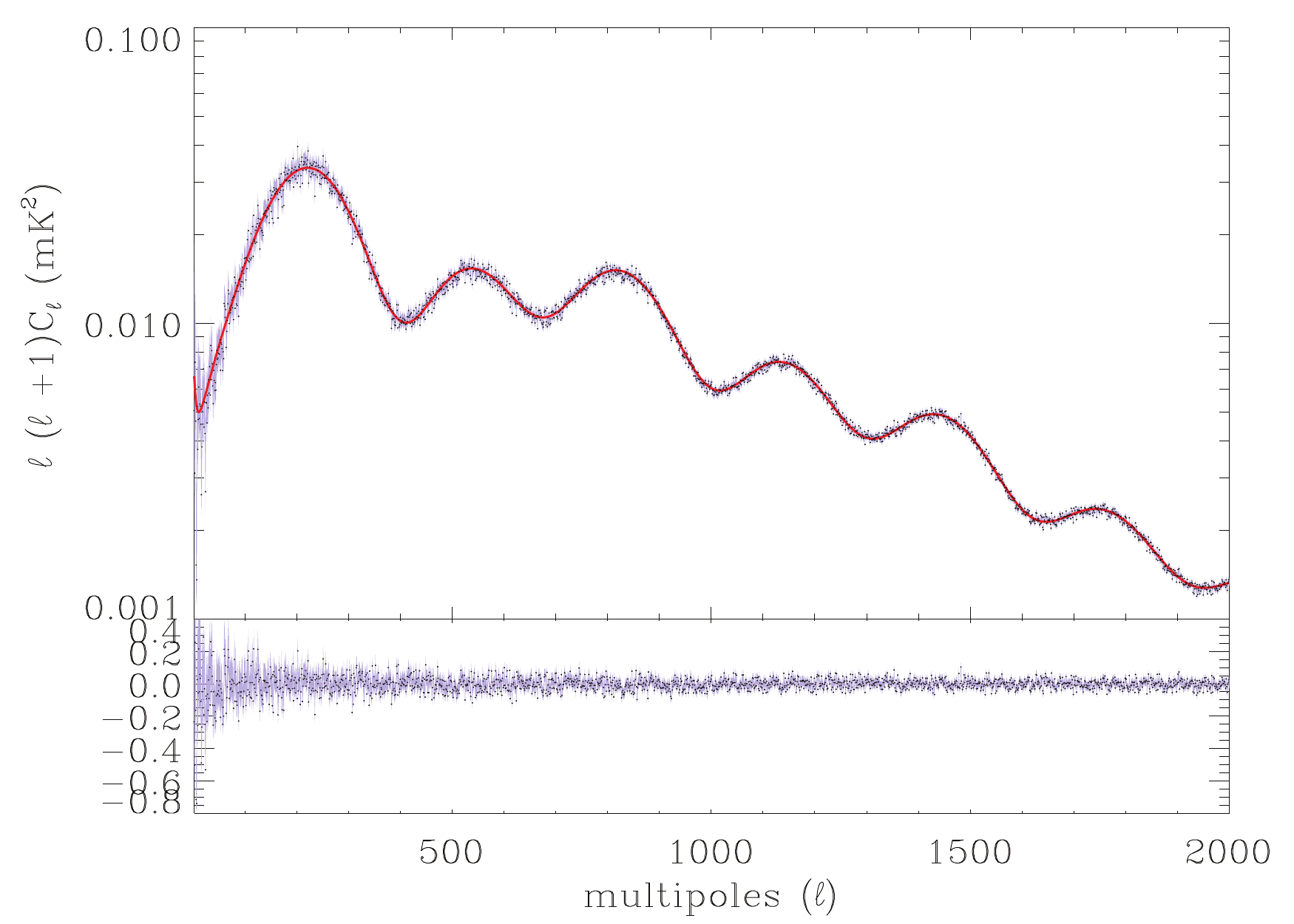}}
\centerline{\includegraphics[scale=0.5]{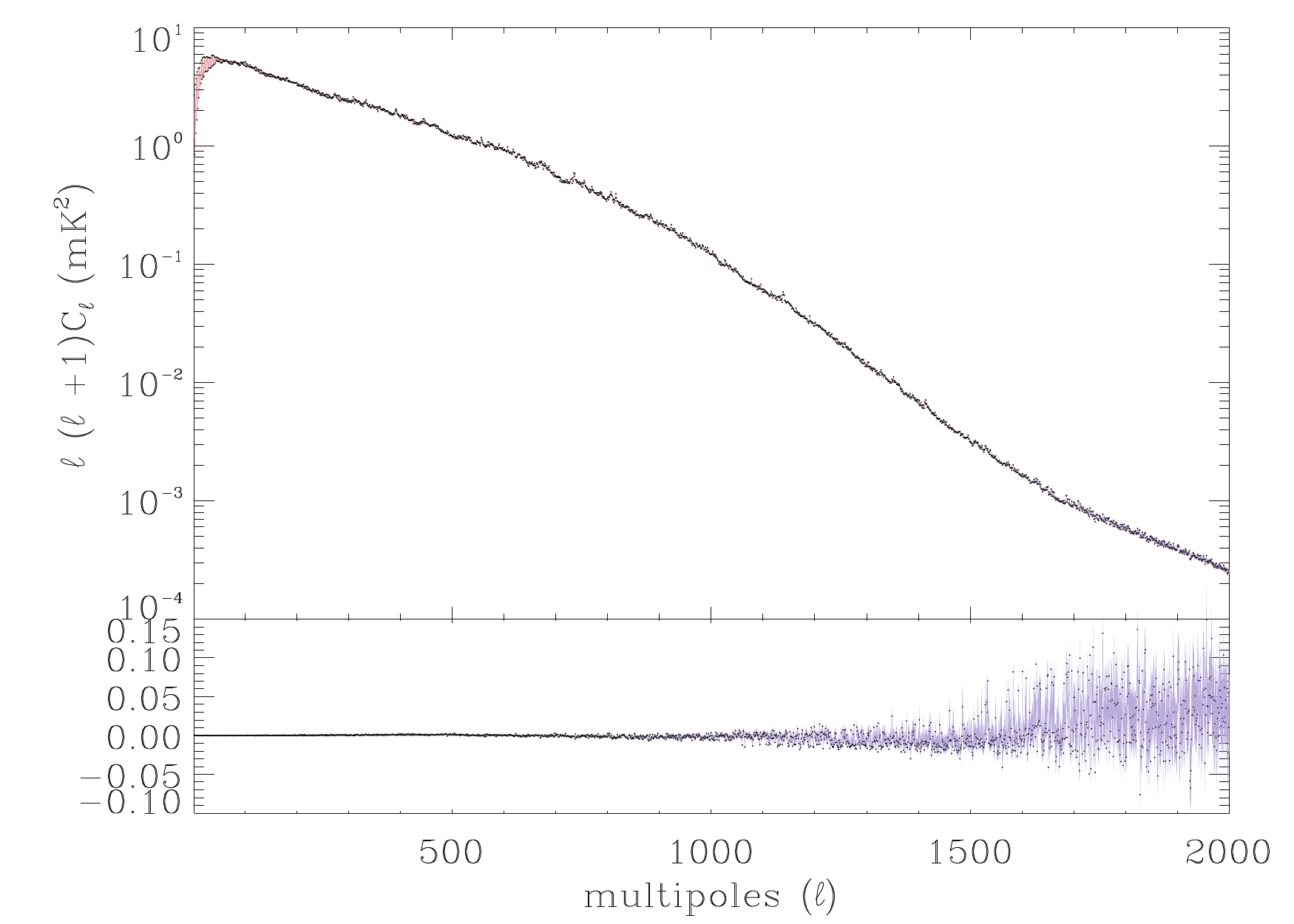}}
\caption{Input and inferred power spectra of the CMB (top) and the sum of the non-CMB components (bottom). In the upper panel of each figure, the black dots  at each multipoles represent the peaks of the marginal posterior, the grey region shows the asymmetric $\pm 1\sigma$-error bar derived from the marginal posterior, the red line is the input power spectrum. The lower panel represents the relative error to the input power spectrum. The sampler accurately recovers the power spectra of the CMB.} \label{fig:c012}
\end{figure}

\begin{figure*}
\centerline{\includegraphics[scale=0.5]{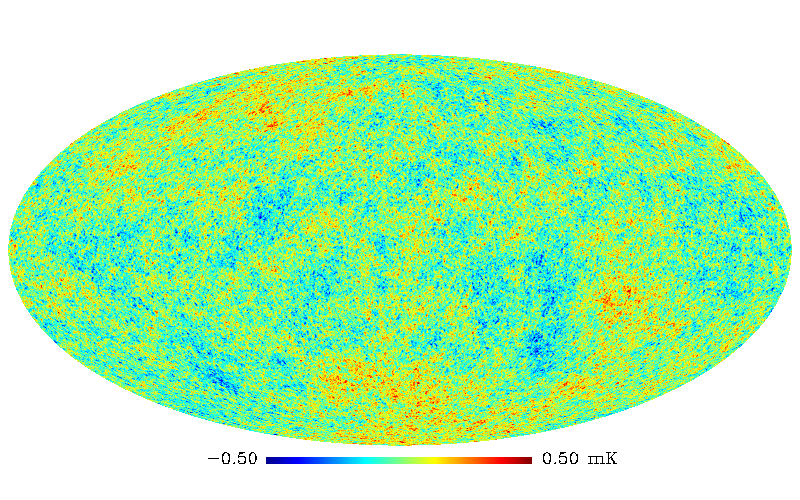}}
\centerline{\includegraphics[scale=0.5]{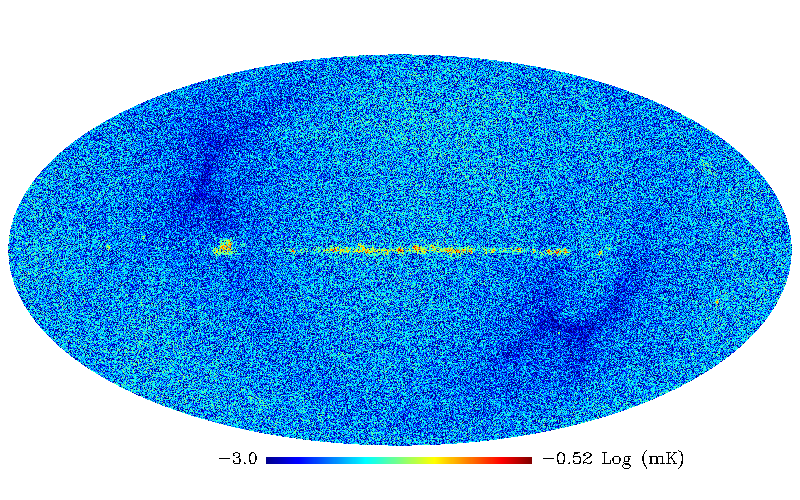}}
\caption{Input and residual CMB map. This residual map represents the absolute value of the difference between the sample mean and the input map. The errors are wider in the galactic plane but the uncertainties in this region of the sky are also larger, as shown on Fig.~\ref{fig:stdmap}. To show the noise in the residual map, it is shown on a decimal logarithm scale.} \label{fig:inout}
\end{figure*}

\begin{figure*}
\centerline{\includegraphics[scale=0.5]{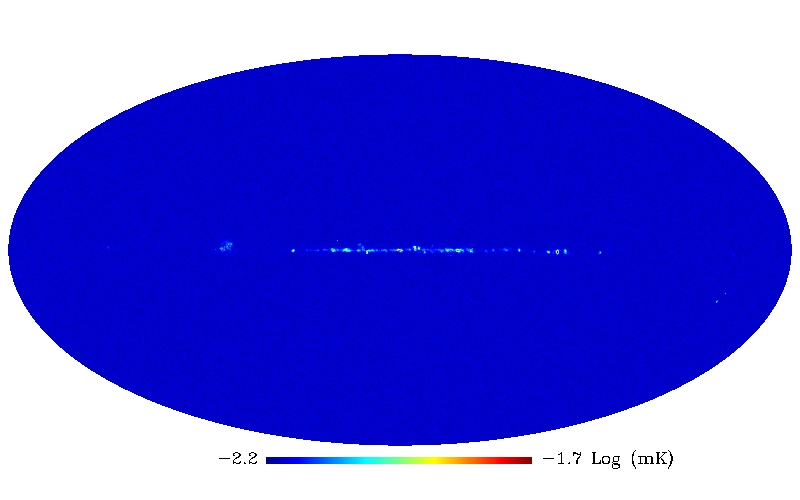}}
\centerline{\includegraphics[scale=0.5]{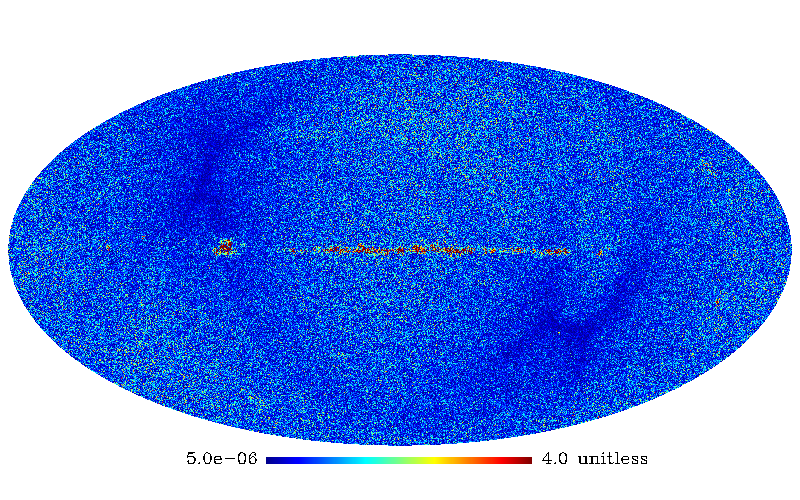}}
\caption{Top: Standard deviation map of the CMB map samples. The posterior distribution is wider in the region of the galactic plane. Bottom: Standardized error map, all red pixel have value 4 or more. This map is the ratio between the residual map of the CMB and its standard deviation map (top). Note that the posterior standard deviation map only represents the part of the uncertainty that is uncorrelated from pixel to pixel while the Bayesian analysis returns a fully correlated error model for the recovered map shown in Fig.~\ref{fig:inout}. Standardizing with the uncorrelated errors reveals two things: the isotropic noise approximation leads to overestimated uncertainties in low noise regions; and an uncorrelated error model does not capture the uncertainties in regions where foregrounds dominate.  See Fig.~\ref{fig:cormap} for a visualization of correlated uncertainties.} \label{fig:stdmap}
\end{figure*}

\begin{figure*}
\centerline{\includegraphics[scale=0.667]{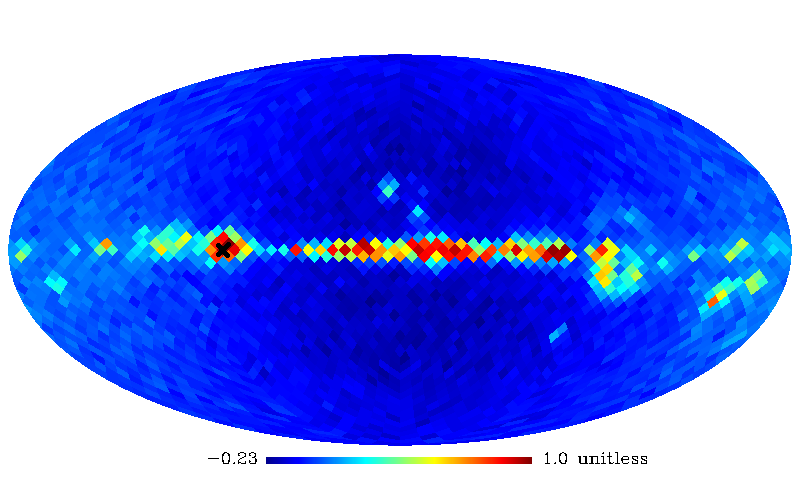}}
\centerline{\includegraphics[scale=0.667]{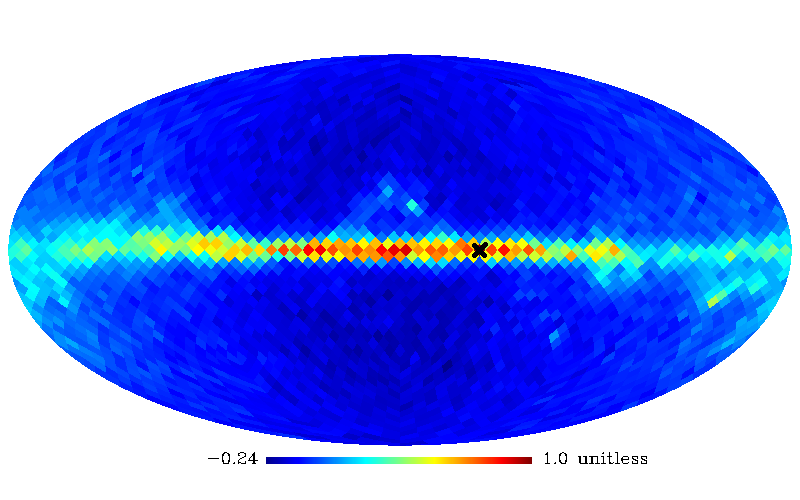}}
\caption{The figure shows two rows of the posterior correlation matrix for the 2 pixels marked by a black cross in each map at HEALPix resolution parameter N$_{\text{side}}=16$. The inferred uncertainties due to foreground removal are highly correlated in the galactic plane and must be taken into account in a meaningful statistical interpretation of the recovered CMB map.}\label{fig:cormap}
\end{figure*}

Considering the power spectra as hyperparameters of the model to be sampled quantifies the error in the prior model of the component maps. Instead of of being static, we let them be constrained by the data, free of any informative prior. De facto, the errors from component separation are encoded in the posterior distribution because the maps and their power spectra are jointly inferred with the mixing matrix. Thus, the PDF over the maps and the power spectra takes into account the systematic errors due to the presence of multiple components in the data and their inference.

\subsection{Power spectrum inference}

All elements of one column of the mixing matrix are fixed to the same arbitrary constant (we chose 1). This prior information leads the sampler to distribute the information contained in the data. Any emission that is constant through frequency is transferred into the power spectrum corresponding to the constant mixing matrix column, and any other emission is transferred into the other power spectra. 
 Since the CMB is the only coherent signal with constant response across all frequencies, our analysis amounts to a CMB power spectrum inference in the presence of foregrounds systematics.

Regarding the other components, the unmixing is not unique. Since there is no physical information on either the power spectra or the frequency spectra, the outcome of these parameters is a mixture of the input parameters that obtains the most probable configuration. \textit{A priori}, we don't expect the individual input power spectra of the non-CMB components to be identifiable as dust and free-free because we force no correlation between the two spatial distributions.\footnote{Physically, dust and free-free are spatially correlated since both of them are more prevalent in the galactic disk than at high latitudes}

Fig.~\ref{fig:c012} shows the inference of the power spectra, the CMB on top and the sum of the non-CMB components below. In order to visualise the inference, we plot the peak of the marginal PDF for each multipole. The peaks are represented by black dots in the upper panel of the each plot of the figure. As expected for the CMB, the shape around the peaks at low multipole is not symmetric. An inverse-gamma distribution fits the individual marginal distributions well, see Fig.~\ref{fig:fit}, as expected (see e.g. \citet{PhysRevD.70.083511}). The grey region represents the shape as if the distribution were a two-sided Gaussian distribution: upper error is one upper standard deviation and the lower error is one lower standard deviation. The solid red line shows the input power spectrum. The lower panel of each plot in the figure shows the relative error to the input power spectrum. The input power spectrum of the CMB lies within the error bars and the recovery is accurate at better than the percent level. For the sum of the non-CMB components, small biases appear from $\ell=1000$. The biases are due to the fact that the correlation between the components are not taken into account, as explain in Sect.~\ref{sec:moch}. These biases are small compared to the CMB power and therefore have no significant effect on the CMB inference.

\subsection{Map inference}

Marginalising the posterior over all parameters but one pixel of one component map leads to a distribution which is consistent with a Gaussian distribution. We therefore consider the sample mean of the map samples, which is an estimate of the mean posterior CMB map, as a reference for a recovered CMB map. Fig.~\ref{fig:inout} shows the input CMB map and the absolute value of the residual map. There is more residual error in the galactic plane because of foreground contamination. Pixels are correlated in the posterior but qualitative errors on the recovered map are given by the sample variance of each pixel.

Fig.~\ref{fig:stdmap} shows a map containing the sample standard deviation of each pixel of the CMB map. As stated above, the errors on the CMB map include the uncertainty due to the presence of galactic emission. We also plot in Fig.~\ref{fig:stdmap} the standardized error map, i.e. the ratio between the residual map and the standard deviation map. The isotropic noise approximation leads to an overestimation of the error bars in the regions of low noise. The per-pixel error is underestimated in highly contaminated regions. The residuals have strong spatial correlations, see Fig.~\ref{fig:cormap}. Another explanation could be that the Gaussian model is too coarse an approximation in regions where the foregrounds are the most intense and highly non-Gaussian. If it is the case, we could use these results to construct masks from the sample variance map to mask the observation maps where necessary, since these regions are correlated to the regions of high variances in the sample variance map (top of Fig.~\ref{fig:stdmap}).

\begin{figure}
\centerline{\includegraphics[scale=0.5]{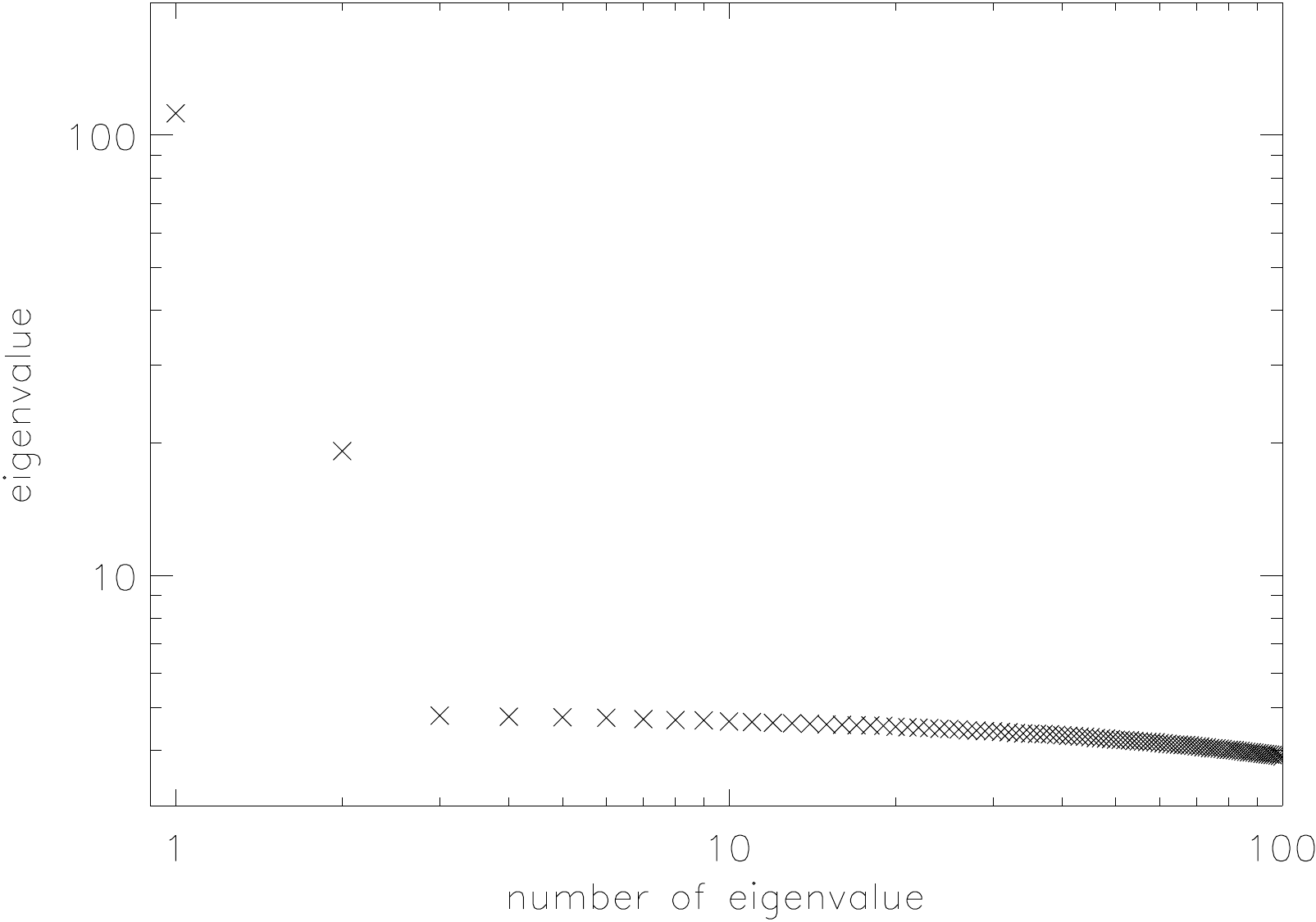}}
\caption{The hundred largest eigenvalues of the posterior correlation matrix of the low resolution CMB map. Two modes dominate.} \label{fig:eigen}
\end{figure}

Pixels in the posterior are correlated and the sample variance map only is not sufficient to fully describe the error on the reconstructed CMB map. To show the correlation, we compute the correlation matrix of the CMB map samples on a lower resolution map. The correlation matrix has a row for every pixel showing the correlation of this pixel to all other pixels. Figure \ref{fig:cormap} shows the correlation maps of two pixels in the galaxy plane. The pixels in the galactic plane are highly correlated, which explains at least part of the excess of $\chi^2$ in the galactic plane of Fig.~\ref{fig:stdmap}. An eigenvalue analysis of the correlation matrix shows that, in addition to noise uncertainties on small scales, the foreground subtraction uncertainties are dominated by a few, highly correlated modes, see Fig.~\ref{fig:eigen}.

\section{Model checking}
\label{sec:moch}

In Fig.~\ref{fig:c012}, the uncertainties on the reconstruction directly rely on the shape of the posterior. Therefore the errors are correctly estimated if the \textit{a priori} model correctly describes the data. It is therefore important to asses the quality of the fit achieved by the model through model checking \citep{gelman1996model}. In order to check for biases in the reconstruction due to assumptions on the statistical properties of the components, we measure the mismatch between the empirical covariance of the data and the covariance of the data model of Eq.~\ref{eq:modelcov}. Following the analysis of \citet{Delabrouille:2002kz}, we examine for each $\ell$ the quantity
\[
D_{\ell} = (2\ell+1)\left[\Tr\left(\hat{R}_{\ell}R_{\ell}^{-1}\right)-\log\left|\hat{R}_{\ell}R_{\ell}^{-1}\right|-\mathrm{N_f}\right]
\]
with $\hat{R}_{\ell}$ the estimation of the data covariance at multipole $\ell$. $D_{\ell}$ is simply the log-likelihood, rewritten as twice the Kullback-Leibler divergence between the two PDF $p_\ell (d\,\vert\,\hat{R_\ell})$ and $p_\ell (d\,\vert\,R_\ell)$ with
\[
p_\ell (d\,\vert\,\Sigma) = \prod_{m=-\ell}^{\ell} \left| 2\pi \Sigma \right|^{-1/2} \exp -\frac{1}{2} d_{\ell m}^T \Sigma^{-1} d_{\ell m} \, .
\]
The posterior PDF on the CMB power spectrum approaches a Gaussian as $\ell$ increases, such that $D_{\ell}$ has the properties of the chi-squared distribution for sufficiently large $\ell$.

The number of degrees of freedom of the distribution is obtained by subtracting the number of stochastic parameters per multipole from the number of degrees of freedom of a symmetric matrix $\mathrm{N_f}\times\mathrm{N_f}$. $\mathrm{N_c}$ spectra are sampled per multipole and the mixing matrix is sampled once for all multipoles. Thus, if the number of degrees of freedom within a mixing matrix is distributed over all multipoles, the number of degrees of freedom of the chi-squared distribution followed by each $D_\ell$ is
\begin{align*}
\mathrm{N_{dof}} &= \mathrm{N_f}\left(\mathrm{N_f}+1\right) / 2 - \left[ \mathrm{N_c} + \frac{(\mathrm{N_f}-1)(\mathrm{N_c}-1)}{(\ell_\mathrm{max}-\ell_\mathrm{min}+1)} \right] \\
 &\simeq \mathrm{N_f}\left(\mathrm{N_f}+1\right) / 2 - \mathrm{N_c} \, .
\end{align*}
In particular $\mathbb{E}\left[D_{\ell}\right] = \mathrm{N_{dof}}$, which does not depend on $\ell$. In our test case $\mathrm{N_{dof}}=7$.

\begin{figure}
\centerline{\includegraphics[scale=0.5]{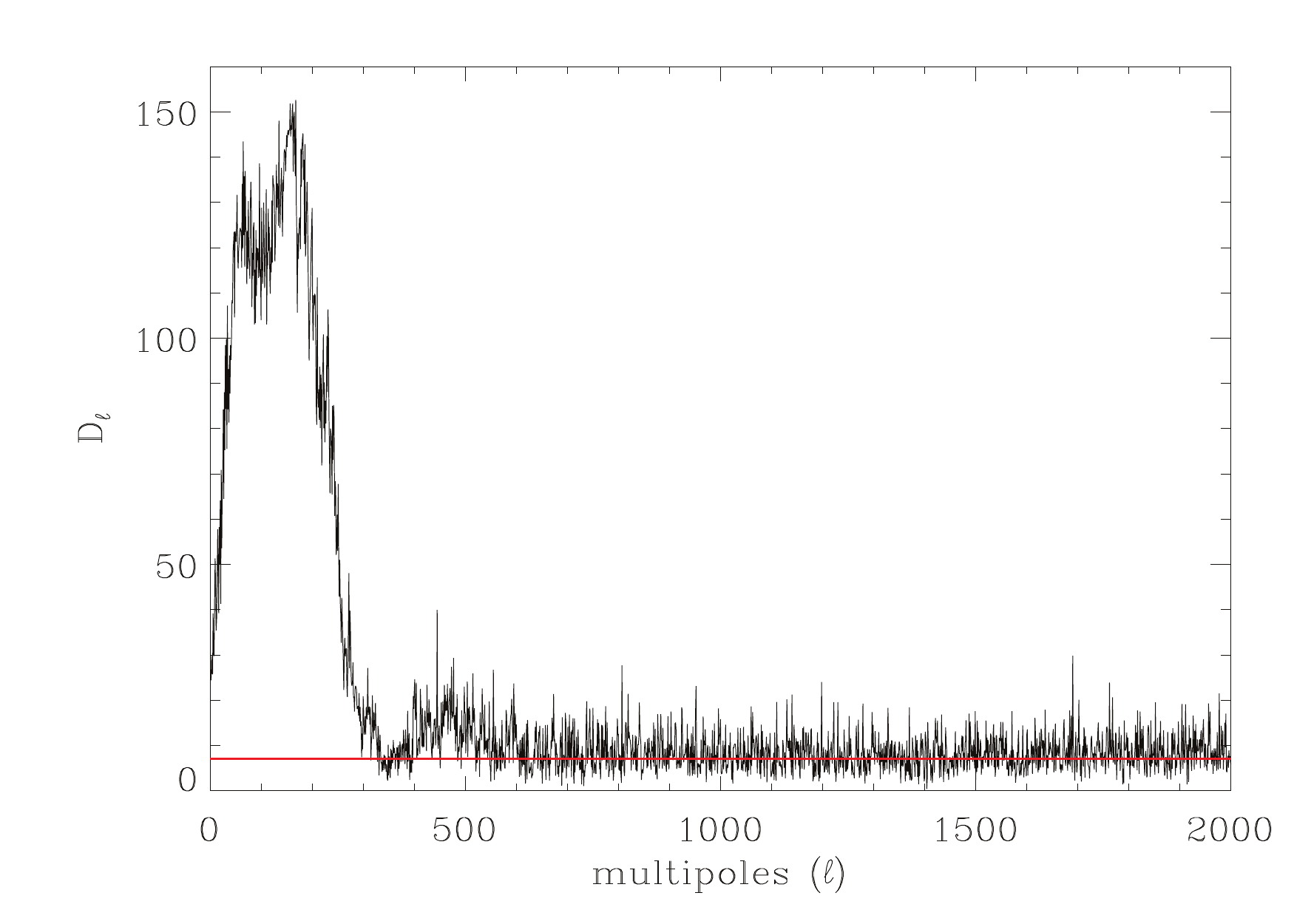}}
\centerline{\includegraphics[scale=0.5]{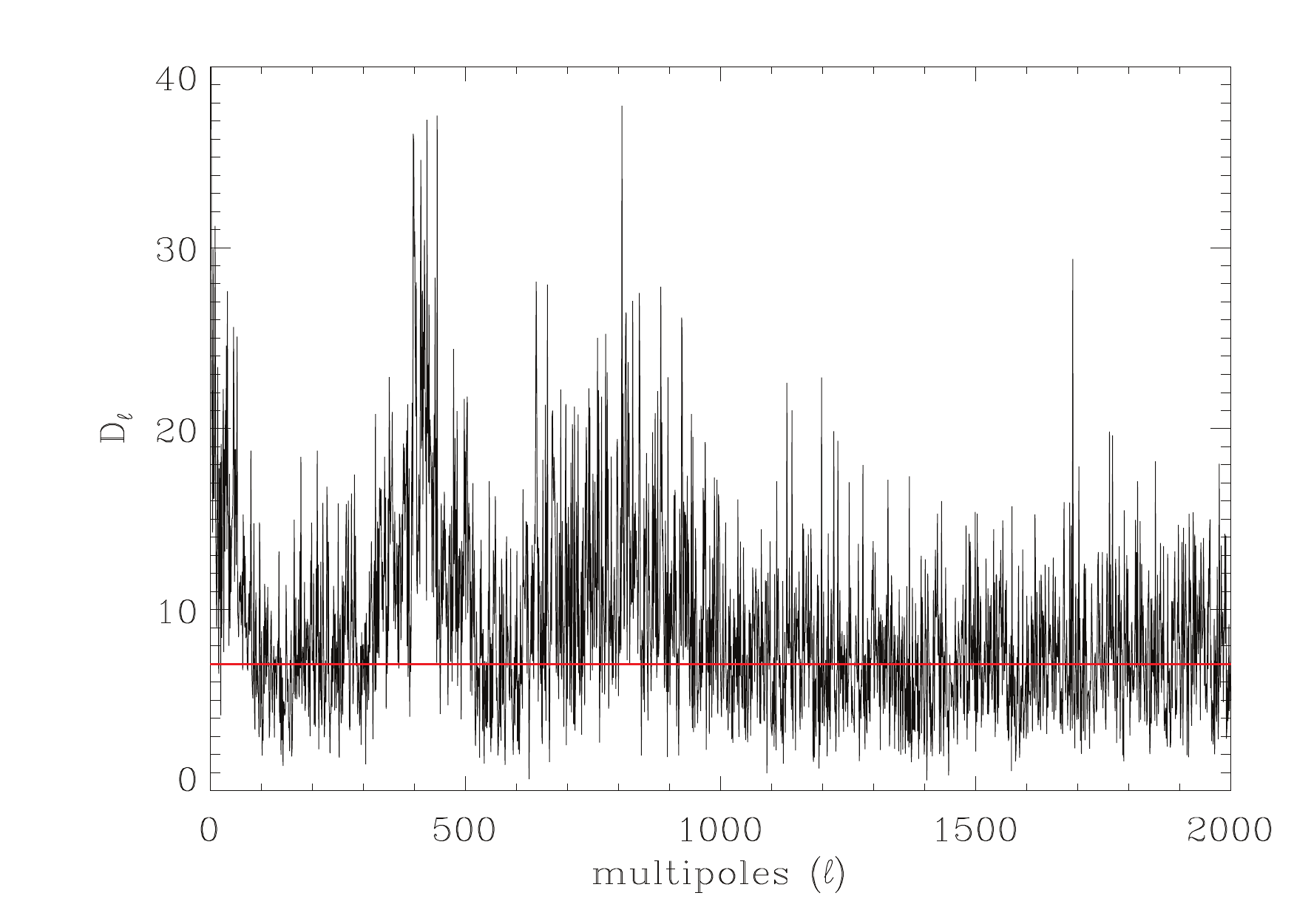}}
\caption{Mismatch between the data and the data modelling. Top - Divergence between the data covariance and input parameters. The large mismatch at low $\ell$ is due to correlations between the input component maps. Bottom - Divergence between the data covariance and the recovered parameters. During sampling, we impose no cross-correlation. Thus the sampler converges towards components that are uncorrelated but whose power spectra are almost capable of capturing the covariance of the input component maps. The red line is the mean of the expected chi-squared distribution that the $D_\ell$'s should approximate.}\label{fig:chis}
\end{figure}

In Fig~\ref{fig:chis} we plot the $D_\ell$'s for $R_{\ell}$ containing the inferred value of $\lbrace A,C \rbrace$. For comparison, we also plot the $D_{\ell}$'s in the case where $A$ and $C$ are set to their input values. Although the input parameters are the true parameters to be recovered, the inferred values have lower mismatch because the components are correlated in the data but not in the model and the sampler finds uncorrelated components that fit the data better. If the foreground modelling matches the statistical properties of the input foregrounds, the $D_{\ell}$'s should follow a chi-squared distribution with a number of degrees of freedom $\mathrm{N_{dof}}$, whose mean $\mathrm{N_{dof}}$ is represented by the horizontal red line on Fig.~\ref{fig:chis}.

\begin{figure}
\centerline{\includegraphics[scale=0.5]{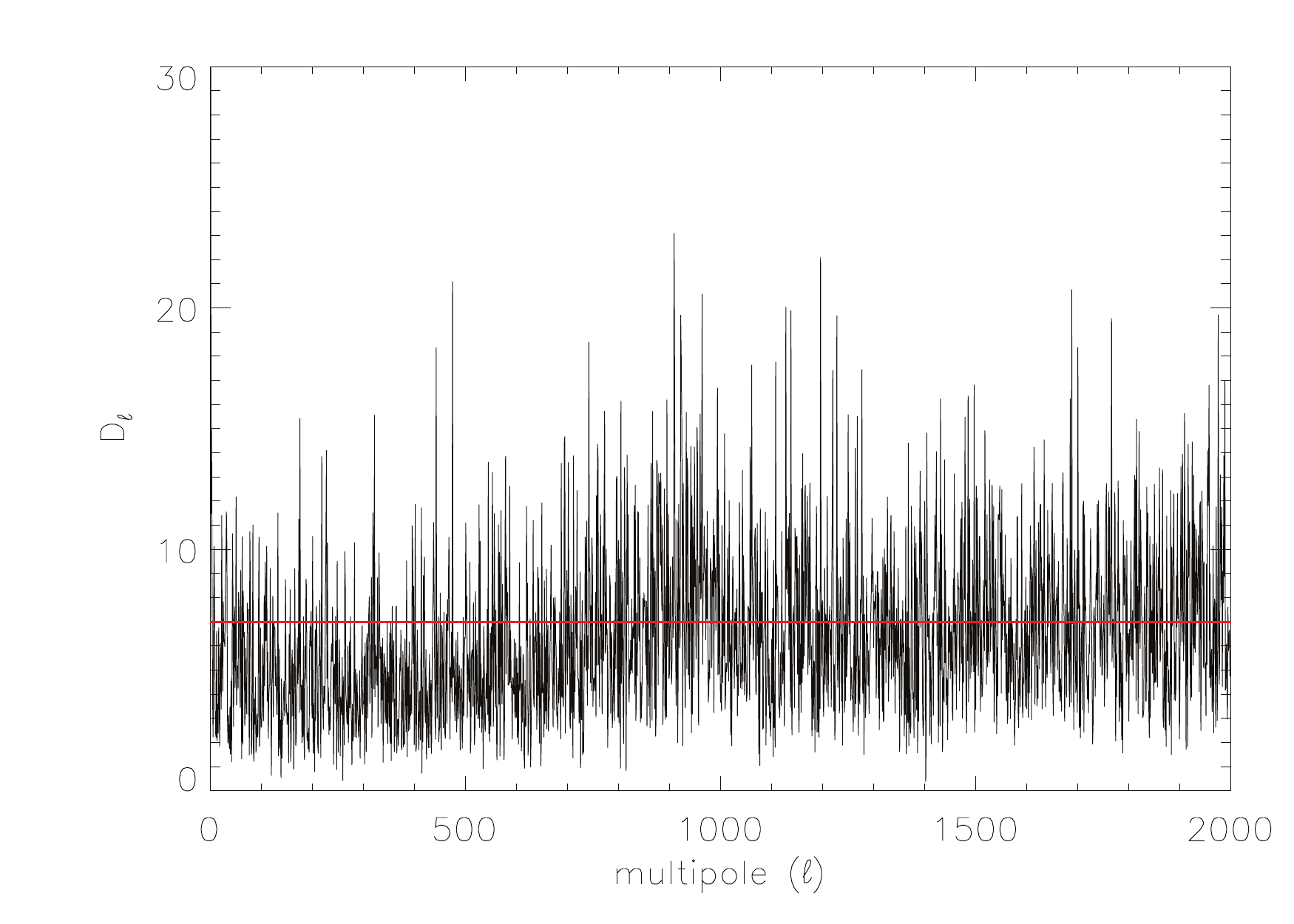}}
\centerline{\includegraphics[scale=0.5]{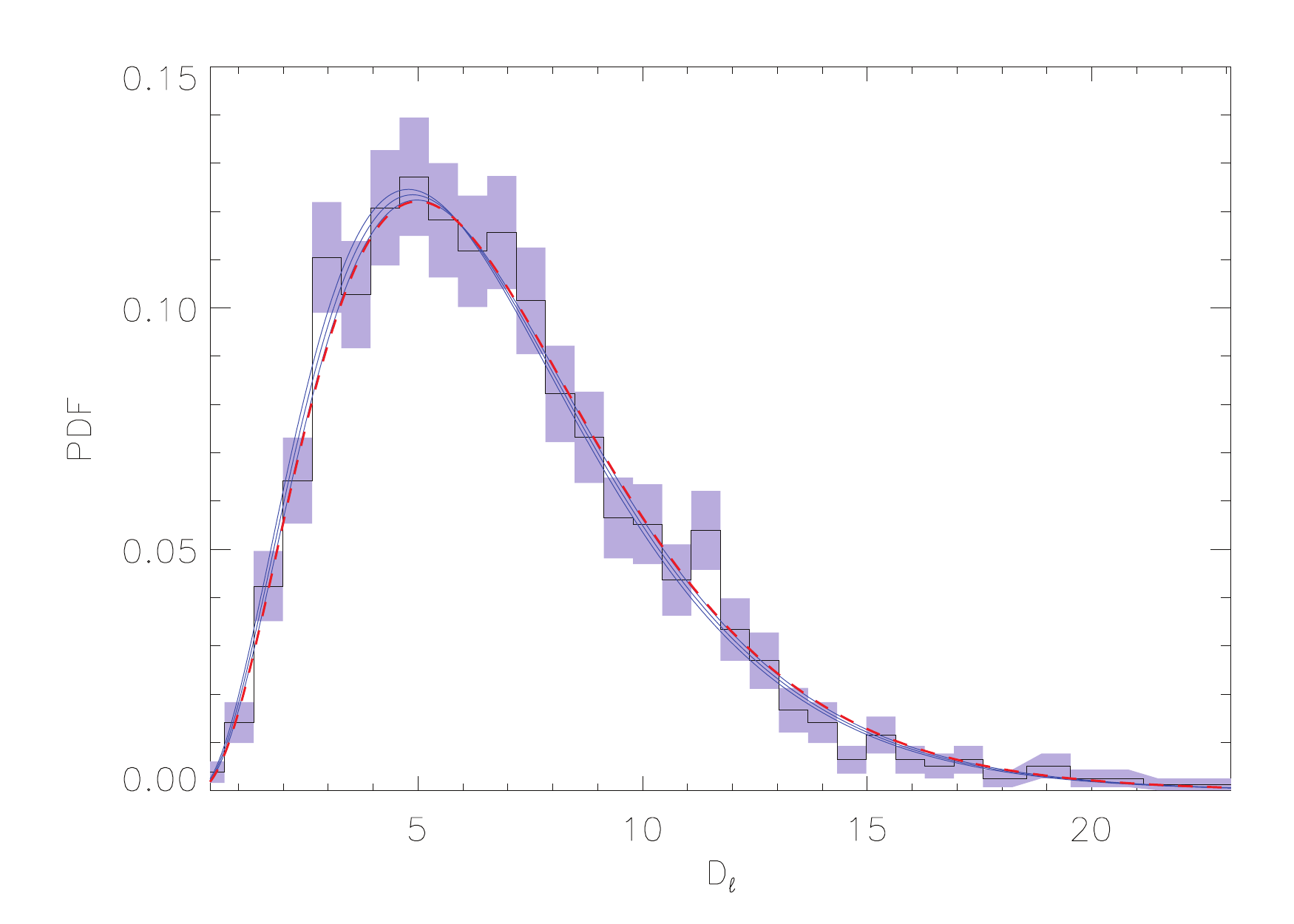}}
\caption{Mismatch between the data and the result of a sampler that includes the input correlations between the components during the sampling process (top) and its PDF (bottom). In the upper panel the red line represents the mean of the expected chi-squared distribution that the $D_{\ell}$'s should follow. In the bottom panel, the blue lines are the best $\chi^2$-fit to a chi-squared curve and the $\pm 1$ standard deviation error from the fit. The red dashed line represents the chi-squared distribution with the expected number of degrees of freedom $\mathrm{N_{dof}} = 7$. The introduction of the correlation between the input component erases the large discrepancy at low multipole.}\label{fig:chiscorr}
\end{figure}

We performed a separation where the cross power spectra of the component are taken into account during the sampling. We do not sample the cross power spectra but we use the covariance of the input components instead, i.e. each $C_{\ell}$ is a non-diagonal matrix but only the diagonal is stochastic. In Fig~\ref{fig:chiscorr} we plot the $D_\ell$'s with the output values of $\lbrace A,C \rbrace$ of such a run. Taking the correlation of the input component maps into account erases the discrepancy at low $\ell$. A chi-squared distribution fits the histogram of the $D_\ell$ for $\ell$ large enough ($\ell > 700$ for this plot). In the figure, the fit is shown by three blue solid lines, the solution of the fit and the $\pm 1 \sigma$ error on the fit. The red dashed line represents the chi-squared distribution with the expected number of degrees of freedom $\mathrm{N_{dof}}$ and it lies within the error bars. The remaining deviations from an $\ell$-independent distribution are due to the differences between the data model and the data actually used. Indeed, the mismatch is completely flat and has the appropriate degree of freedom if a set of simulations completely consistent with the data model is used. Also, taking the correlations between components into account erases the biases in the inference of the power spectrum of the sum of the components.

Considering a Gaussian model for the components and neglecting the correlations between the components do not affect the reconstruction of the CMB power spectrum since the CMB is not correlated to the foregrounds. On another hand, neglecting the correlation between the non-CMB components is too coarse an approximation if one also wants to recover the galactic components.

\section{Comparison to previous component separation methods}
\label{sec:prevmeth}

\subsection{Comparison to Commander}

As our method, Commander \citep{Eriksen:2005dr,Eriksen:2007mx} is a Bayesian formulation of the joint component separation and CMB power spectrum inference problem. The main difference between the two approaches is the parametrisation of the problem.

Commander makes intensive use of parametric models to describe the physical emissions while we adopt a quasi-blind and a phenomenological description of the different components. Thus Commander infers maps and CMB power spectrum within a constraining physical model, and therefore the most probable values of the parameters and their errors depends on this model. We do not rely on any physical assumption, except for the constant response of CMB signal across frequencies. The drawback is that only the recovered map and power spectrum of the CMB have a physical meaning but the non-trivial difference is that our results do not depend on physical modelling assumptions.

In addition, Commander works at the map level whereas our method works at the multipoles level. Thus, our sampler is fast and allows treatment of high resolution data maps.

\subsection{Comparison to SMICA}

Spectral Matching Independent Component Analysis (SMICA) \citep{2008arXiv0803.1814C} is a method that provides component power spectra and mixing coefficients. The parameters are estimated by finding the best match between an $A$- and $C$-dependent covariance and an empirical data covariance. Depending on the binning of the power spectrum, and following the original formulation of SMICA, the method is equivalent to a maximisation with respect to $\lbrace A,C\rbrace$ of the SMICA likelihood of Eq.~\ref{eq:smica}. One among several applications of this method is to apply a Wiener filter to the data with the solution of the maximisation in order to recover the component maps.

We can understand SMICA analysis from a Bayesian perspective as follows:
\begin{enumerate}
\item Begin with the Bayesian formulation, Eq~\ref{eq:bayesTh},
\item Choose Gaussian priors for the components, flat priors for the mixing matrix and the power spectra,
\item Marginalise the posterior over all component maps,
\item Maximise the obtained marginal distribution with respect to $A$ and $C$.
\end{enumerate}
In the Bayesian formulation, maximising the marginal posterior $P( A,C \,\vert\,d)$ in Eq.~\ref{eq:margin} is equivalent to maximising the likelihood in Eq.~\ref{eq:smica}. Thus, instead of just finding the peak of the distribution like SMICA does, the Bayesian sampler explores the whole distribution over $A$ and $C$, and therefore returns an error model for the cleaned CMB map that includes foreground cleaning uncertainties. In addition, the power spectrum inference is marginalised over foreground cleaning uncertainties.

\subsection{Comparison to standard ILC}

The Internal Linear Combination (ILC) method \citep{Bennett:1992} provides a map of a component, given its frequency dependence. The method involves a weighted average of the observation maps in order to cancel all components but the one of interest. Usually, ILC is derived by minimizing the variance of a linear combination of the observation maps. The remaining fluctuations are the CMB anisotropies.

ILC can also be derived by adopting a Bayesian point of view:
\begin{enumerate}
\item Begin with the Bayesian formulation, Eq~\ref{eq:bayesTh},
\item Choose Gaussian priors for the component, flat priors for the mixing matrix and the power spectra,
\item Marginalise the posterior over all component maps but the CMB map,
\item Maximise the obtained marginal distribution with regard to the CMB map.
\end{enumerate}
The solution is
\begin{equation}
\label{eq:sol}
\hat{s}_{\mathrm{cmb},\ell m} = \frac{e^T \left( N_{\ell} + A C_{\ell} A^T \right)^{-1}}{C_{\mathrm{cmb},\ell}^{-1}+e^T \left( N_{\ell} + A C_{\ell} A^T \right)^{-1} e} d_{\ell m}
\end{equation}
where here the $\mathrm{N_f}\times(\mathrm{N_c-1})$ mixing matrix $A$ has no column dedicated to the CMB, $C_{\ell}$ is the covariance of all components but the CMB at multipole $\ell$ and $e=(1\dots1)^T$, i.e. the frequency response of the CMB. The standard ILC formula is
\begin{equation}
\label{eq:ilc}
\hat{s}_{\mathrm{cmb},\ell m} = \frac{e^TC_d^{-1}}{e^TC_d^{-1}e} d_{\ell m}
\end{equation}
where $C_d$ is an estimate of the data covariance
\begin{equation}
C_{d_{\ell m}} = C_{\mathrm{cmb}, \ell} ee^T + A_{\mathrm{true}} C_{\mathrm{true},\ell m} A_{\mathrm{true}}^T + N_{\ell m} \, .
\end{equation}
If the CMB fluctuations are neglected in the ILC formula, and if the prior variance of the CMB is infinite (i.e. flat prior) in the Bayesian formula~\ref{eq:sol} then the two approaches are equivalent. The ILC method naturally chooses an estimation of the true mixing matrix and the true power spectra of the components. That is why, despite its very simple formulation and its approximations, the ILC method is physically consistent.

\section{Discussion and future works}
\label{sec:discussion}
The fact that the sampler can not distinguish between non-CMB emissions is due to a additional degrees of freedom. Without any information about the physical emissions, all the recovered components others than CMB are a mixture of the true signals. Putting a prior either on the mixing matrix, i.e. import knowledge about the frequency spectra, or on the shapes of the component power spectra would break the degeneracies. Although the constraints of the priors can be controlled, the blindness of the method would be lost with this introduction of \textit{a priori} information. Furthermore, in this paper we impose decorrelation between the component maps. The foregrounds could be modelled in more detail to get a full component separation method, but the focus here is on CMB reconstruction.

The next step is to apply the method to real data. The main problem is the instrumental noise. In this paper we assume a simple noise model. To deal with real data noise requires developing a more realistic noise model. A full model of correlated noise involves very large covariance matrices. Therefore alternative ways to deal with noise like half-ring half difference maps or noise simulations should be considered.

Unresolved point sources appear as extra power at small angular scales of the inferred CMB power spectrum. Masking the listed point sources and inpainting in the mask would be a way to address the point sources issue. In addition allowing the components to mix differently in different regions of the sky or in different angular scale ranges by allocating different mixing matrices in each range would reduce mismatch due to lack of coherence. Also, foregrounds that are not fully coherent from frequency to frequency may be modelled by increasing the number of components in our model.

In this work we reconstructed the CMB maps on the full sky. It remains to be seen if this is achievable on realistic data. Since our approach is similar to the one by SMICA, which provides a clean map on a large fraction of the sky in Planck analysis \citep{PlanckXII}, treatment of almost full sky data should be feasible. Since we work under the assumption of diagonal covariances in $\ell$-space the effect of a small mask needs to be tested. It may be possible to avoid masking using inpainting. If necessary a Wiener filter method such as \citet{Elsner:2012fe} could be used to implement a full Gibbs sampling approach.

If joint modelling of foregrounds allows working with a large part of the sky, we may be able to ignore mode coupling effects due to the mask with high accuracy.

Although in this paper we chose to address component separation with a blind analysis, we can use the flexibility of the Bayesian formalism in order to introduce physical parametrisation of the problem. The current understanding of physical galactic and extra-galactic phenomena can be progressively introduced by a more detailed data model and by assigning a parametrized prior PDF on the foregrounds. The CMB power spectrum is also highly parametrizable since its shape depends on a small number of cosmological parameters \citep{PlanckCollaboration+13_cosmopars}. Thus a joint inference of cleaned CMB map, CMB power spectrum and cosmological parameters is conceivable, thanks to fast and accurate Boltzmann code emulators like PICO \citep{2007ApJ...654....2F}. A less parametric approach would be to exploit the smoothness of the power spectra through binning or representation in terms of smooth basis function, such as splines.

Also, as in other component separation methods, the Bayesian formulation presented in this paper can be extended to infer the CMB polarisation power spectrum.

We leave these further explorations to future work.

\section{Conclusion}
\label{sec:conclusion}
We have presented a new formulation for the CMB foreground cleaning. In our analysis, we avoid physical parametrisations, except that the CMB behaves like a black body and we model the components as Gaussian random fields. The CMB is then cleaned by jointly inferring CMB, galactic residuals, and point source power spectra and frequency spectra. This Bayesian method provides an evaluation of a posterior PDF for the CMB power spectrum which thus takes into account uncertainties due to the removal of foreground contamination. Full maps of CMB anisotropies are recovered with their own PDF which reveals that the dominant foreground residuals are captured in terms of a small number of error modes. We also showed that previous component separation methods can be derived as special cases of our Bayesian formulation, which thus provides a unified approach for semi-blind foreground cleaning from multi-frequency CMB data.

\begin{acknowledgements}
FV thanks Franz Elsner for useful discussion in the early stages of the work. This work made in the ILP LABEX (under reference ANR-10-LABX-63) was supported by French state funds managed by the ANR within the Investissements d'Avenir program under reference ANR-11-IDEX-0004-02. This work was also partially supported by NSF AST 07-08849. B.D.W. is supported by a senior Excellence Chair by the Agence Nationale de Recherche (ANR-10-CEXC-004-01) and a Chaire Internationale at the Universitée Pierre et Marie Curie
\end{acknowledgements}

\bibliographystyle{aa}
\bibliography{biblatex}

\begin{thebibliography}{37}
\expandafter\ifx\csname natexlab\endcsname\relax\def\natexlab#1{#1}\fi

\bibitem[{André {et~al.}(2014)}]{Andre:2013nfa}
André, P. {et~al.} 2014, JCAP, 1402, 006

\bibitem[{Baccigalupi {et~al.}(2000)Baccigalupi, Bedini, Burigana, De~Zotti,
  Farusi, {et~al.}}]{Baccigalupi:2000xy}
Baccigalupi, C., Bedini, L., Burigana, C., {et~al.} 2000, Mon. Not. Roy.
  Astron. Soc., 318, 769

\bibitem[{Baumann {et~al.}(2009)}]{Baumann:2008aq}
Baumann, D. {et~al.} 2009, AIP Conf. Proc., 1141, 10

\bibitem[{Bennett {et~al.}(1992)Bennett, Smoot, Hinshaw, Wright, Kogut, \&
  de~Amici}]{Bennett:1992}
Bennett, C., Smoot, G., Hinshaw, G., {et~al.} 1992, Astrophys. J., 396, L7

\bibitem[{{Bennett} {et~al.}(2012){Bennett}, {Larson}, {Weiland}, {Jarosik},
  {Hinshaw}, {Odegard}, {Smith}, {Hill}, {Gold}, {Halpern}, {Komatsu}, {Nolta},
  {Page}, {Spergel}, {Wollack}, {Dunkley}, {Kogut}, {Limon}, {Meyer}, {Tucker},
  \& {Wright}}]{2012arXiv1212.5225B}
{Bennett}, C.~L., {Larson}, D., {Weiland}, J.~L., {et~al.} 2012,
  arXiv:1212.5225

\bibitem[{Bouchet {et~al.}(2011)}]{Bouchet:2011ck}
Bouchet, F.~R. {et~al.} 2011, arXiv:1102.2181

\bibitem[{Cardoso(1998)}]{Cardoso:1998}
Cardoso, J.-F. 1998, Proceedings of the IEEE, 86, 2009

\bibitem[{{Cardoso} {et~al.}(2008){Cardoso}, {Martin}, {Delabrouille},
  {Betoule}, \& {Patanchon}}]{2008arXiv0803.1814C}
{Cardoso}, J.-F., {Martin}, M., {Delabrouille}, J., {Betoule}, M., \&
  {Patanchon}, G. 2008, arXiv:0803.1814

\bibitem[{{Cardoso} {et~al.}(2002){Cardoso}, {Snoussi}, {Delabrouille}, \&
  {Patanchon}}]{2002astro.ph..9466C}
{Cardoso}, J.-F., {Snoussi}, H., {Delabrouille}, J., \& {Patanchon}, G. 2002,
  arXiv:astro-ph/0209466

\bibitem[{Delabrouille {et~al.}(2012)Delabrouille, Betoule, Melin,
  Miville-Deschenes, Gonzalez-Nuevo, {et~al.}}]{Delabrouille:2012ye}
Delabrouille, J., Betoule, M., Melin, J.-B., {et~al.} 2012, arXiv:1207.3675

\bibitem[{Delabrouille {et~al.}(2008)Delabrouille, Cardoso, Jeune, Betoule,
  Fay, {et~al.}}]{Delabrouille:2008qd}
Delabrouille, J., Cardoso, J.-F., Jeune, M.~L., {et~al.} 2008

\bibitem[{Delabrouille {et~al.}(2003)Delabrouille, Cardoso, \&
  Patanchon}]{Delabrouille:2002kz}
Delabrouille, J., Cardoso, J.-F., \& Patanchon, G. 2003, Mon. Not. Roy. Astron.
  Soc., 346, 1089

\bibitem[{{Dickinson} {et~al.}(2003){Dickinson}, {Davies}, \&
  {Davis}}]{2003MNRAS.341..369D}
{Dickinson}, C., {Davies}, R.~D., \& {Davis}, R.~J. 2003, \mnras, 341, 369

\bibitem[{Elsner \& Wandelt(2013)}]{Elsner:2012fe}
Elsner, F. \& Wandelt, B.~D. 2013, Astron. Astrophys., 549, A111

\bibitem[{Eriksen {et~al.}(2008)Eriksen, Jewell, Dickinson, Banday, Gorski,
  {et~al.}}]{Eriksen:2007mx}
Eriksen, H., Jewell, J., Dickinson, C., {et~al.} 2008, Astrophys. J., 676, 10

\bibitem[{Eriksen {et~al.}(2006)Eriksen, Dickinson, Lawrence, Baccigalupi,
  Banday, {et~al.}}]{Eriksen:2005dr}
Eriksen, H.~K., Dickinson, C., Lawrence, C., {et~al.} 2006, Astrophys. J., 641,
  665

\bibitem[{{Fendt} \& {Wandelt}(2007)}]{2007ApJ...654....2F}
{Fendt}, W.~A. \& {Wandelt}, B.~D. 2007, \apj, 654, 2

\bibitem[{Fernandez-Cobos {et~al.}(2012)Fernandez-Cobos, Vielva, Barreiro, \&
  Martinez-Gonzalez}]{FernandezCobos:2011bm}
Fernandez-Cobos, R., Vielva, P., Barreiro, R., \& Martinez-Gonzalez, E. 2012,
  \mnras, 3, 2162

\bibitem[{Gelman \& Meng(1996)}]{gelman1996model}
Gelman, A. \& Meng, X.-L. 1996, Model checking and model improvement
  (Springer), 189--201

\bibitem[{{G{\'o}rski} {et~al.}(2005){G{\'o}rski}, {Hivon}, {Banday},
  {Wandelt}, {Hansen}, {Reinecke}, \& {Bartelmann}}]{2005ApJ...622..759G}
{G{\'o}rski}, K.~M., {Hivon}, E., {Banday}, A.~J., {et~al.} 2005, \apj, 622,
  759

\bibitem[{{Gratton}(2008)}]{2008arXiv0805.0093G}
{Gratton}, S. 2008, arXiv:0805.0093

\bibitem[{{Jarosik} {et~al.}(2011){Jarosik}, {Bennett}, {Dunkley}, {Gold},
  {Greason}, {Halpern}, {Hill}, {Hinshaw}, {Kogut}, {Komatsu}, {Larson},
  {Limon}, {Meyer}, {Nolta}, {Odegard}, {Page}, {Smith}, {Spergel}, {Tucker},
  {Weiland}, {Wollack}, \& {Wright}}]{2011ApJS..192...14J}
{Jarosik}, N., {Bennett}, C.~L., {Dunkley}, J., {et~al.} 2011, \apjs, 192, 14

\bibitem[{Jungman {et~al.}(1996)Jungman, Kamionkowski, Kosowsky, \&
  Spergel}]{Jungman:1995bz}
Jungman, G., Kamionkowski, M., Kosowsky, A., \& Spergel, D.~N. 1996, Phys.
  Rev., D54, 1332

\bibitem[{Lewis {et~al.}(2000)Lewis, Challinor, \& Lasenby}]{Lewis:1999bs}
Lewis, A., Challinor, A., \& Lasenby, A. 2000, Astrophys. J., 538, 473

\bibitem[{Maino {et~al.}(2001)Maino, Farusi, Baccigalupi, Perrotta, Banday,
  {et~al.}}]{Maino:2001vz}
Maino, D., Farusi, A., Baccigalupi, C., {et~al.} 2001, arXiv:astro-ph/0108362

\bibitem[{Moudden {et~al.}(2005)Moudden, Cardoso, Starck, \&
  Delabrouille}]{Moudden:2004wi}
Moudden, Y., Cardoso, J.-F., Starck, J.-L., \& Delabrouille, J. 2005, EURASIP
  J. Appl. Signal Process., 15, 2437

\bibitem[{Mukhanov(2013)}]{Mukhanov:2013tua}
Mukhanov, V. 2013, arXiv:1303.3925

\bibitem[{{Planck Collaboration} {et~al.}(2013{\natexlab{a}}){Planck
  Collaboration}, {Ade}, {Aghanim}, {Alves}, {Armitage-Caplan}, {Arnaud},
  {Ashdown}, {Atrio-Barandela}, {Aumont}, {Aussel}, \& et~al.}]{PlanckI}
{Planck Collaboration}, {Ade}, P.~A.~R., {Aghanim}, N., {et~al.}
  2013{\natexlab{a}}, arXiv:13035062

\bibitem[{{Planck Collaboration} {et~al.}(2013{\natexlab{b}}){Planck
  Collaboration}, {Ade}, {Aghanim}, {Armitage-Caplan}, {Arnaud}, {Ashdown},
  {Atrio-Barandela}, {Aumont}, {Baccigalupi}, {Banday}, \& et~al.}]{PlanckXII}
{Planck Collaboration}, {Ade}, P.~A.~R., {Aghanim}, N., {et~al.}
  2013{\natexlab{b}}, arXiv:1303.5072

\bibitem[{{Planck Collaboration} {et~al.}(2013{\natexlab{c}}){Planck
  Collaboration}, {Ade}, {Aghanim}, {Armitage-Caplan}, {Arnaud}, {Ashdown},
  {Atrio-Barandela}, {Aumont}, {Baccigalupi}, {Banday}, \&
  et~al.}]{PlanckCollaboration+13_cosmopars}
{Planck Collaboration}, {Ade}, P.~A.~R., {Aghanim}, N., {et~al.}
  2013{\natexlab{c}}, arXiv:1303.5076

\bibitem[{{Planck Collaboration} {et~al.}(2013{\natexlab{d}}){Planck
  Collaboration}, {Aghanim}, {Armitage-Caplan}, {Arnaud}, {Ashdown},
  {Atrio-Barandela}, {Aumont}, {Baccigalupi}, {Banday}, {Barreiro}, \&
  et~al.}]{PlanckIII}
{Planck Collaboration}, {Aghanim}, N., {Armitage-Caplan}, C., {et~al.}
  2013{\natexlab{d}}, arXiv:1303.5064

\bibitem[{Schaffer {et~al.}(2011)Schaffer, Crawford, Aird, Benson, Bleem,
  {et~al.}}]{Schaffer:2011mz}
Schaffer, K., Crawford, T., Aird, K., {et~al.} 2011, Astrophys. J., 743, 90

\bibitem[{Sehgal {et~al.}(2010)Sehgal, Bode, Das, Hernandez-Monteagudo,
  Huffenberger, {et~al.}}]{Sehgal:2009xv}
Sehgal, N., Bode, P., Das, S., {et~al.} 2010, Astrophys. J., 709, 920

\bibitem[{{Starck} {et~al.}(2013){Starck}, {Donoho}, {Fadili}, \&
  {Rassat}}]{2013A&A...552A.133S}
{Starck}, J.-L., {Donoho}, D.~L., {Fadili}, M.~J., \& {Rassat}, A. 2013, \aap,
  552, A133

\bibitem[{Starck {et~al.}(2004)Starck, Elad, \& Donoho}]{Starck:896984}
Starck, J.-L., Elad, M., \& Donoho, D.~L. 2004, Adv. Imag. Elect. Phys., 132,
  287

\bibitem[{Tegmark(1997)}]{Tegmark:1997zz}
Tegmark, M. 1997, arXiv:astro-ph/9712038

\bibitem[{Wandelt {et~al.}(2004)Wandelt, Larson, \&
  Lakshminarayanan}]{PhysRevD.70.083511}
Wandelt, B.~D., Larson, D.~L., \& Lakshminarayanan, A. 2004, Phys. Rev. D, 70,
  083511

\end{thebibliography}

\end{document}